\documentclass[twocolumn,floatfix,prb,aps,showpacs]{revtex4-2}
\usepackage{graphicx,amsmath,amssymb,color,nicefrac,multirow, makecell, ragged2e,float}
\usepackage[unicode=true,colorlinks=true,linkcolor=blue,citecolor=blue,urlcolor=blue]{hyperref}
\usepackage[titletoc,title]{appendix}

\newcommand{\be}{\begin{equation}}
\newcommand{\ee}{\end{equation}}

\newcommand{\ba}{\begin{eqnarray}}
\newcommand{\ea}{\end{eqnarray}}

\def\beq{\begin{eqnarray}}
\def\eeq{\end{eqnarray}}

\makeatletter
\newcommand*{\rom}[1]{\expandafter\@slowromancap\romannumeral #1@}
\makeatother

\newcommand{\non}{\nonumber\\}
\usepackage{bm,physics,enumerate,booktabs,ulem}

\begin{document}
\title{Superconductivity in a system of interacting spinful semions}
\author{Koji Kudo and Jonathan Schirmer}
\affiliation{The Pennsylvania State University, 104 Davey Lab, University Park, Pennsylvania 16802, USA}
\begin{abstract}
 Non-interacting particles obeying certain fractional statistics have been 
 predicted to exhibit superconductivity. We discuss the issue in an 
 attractively interacting system of spinful semions on a lattice by numerically
 investigating the presence of off-diagonal long-range order at zero temperature. For this purpose,
 we construct a Hubbard model wherein two semions with opposite spin can 
 virtually coincide while maintaining consistency with the fractional braiding 
 statistics. Clear off-diagonal long range order is seen in the strong coupling
 limit, consistent with the expectation that a pair of semions obeys Bose 
 statistics. We find that the semion system behaves similarly to a system of fermions with the same attractive Hubbard $U$ interaction for a wide range of $U$, suggesting that semions also undergo a BCS to BEC crossover
as a function of $U$.
\end{abstract}

\maketitle

\section{Introduction}
The topology of configuration space determines the possible quantum statistics of a system of identical particles~\cite{Wu84b}. In two
dimensions, the fundamental group of the many-particle configuration space is 
the braid group, which allows exotic particles, namely anyons~\cite{Wilczek82},
that obey statistics beyond bosons and fermions. The emergence of anyons as
elementary excitations is a defining feature of the so-called topological 
order~\cite{Wen95}, which has demonstrated how topology enriches 
phases of matter beyond Landau's symmetry-breaking paradigm.
A typical example of topologically ordered phases is the fractional quantum
Hall effect~\cite{Tsui82}. Special properties such as (non-Abelian) anyon 
excitations~\cite{Arovas84,Moore91,Wen91}, 
fractional charges~\cite{Laughlin83}, and the topological 
degeneracy~\cite{Haldane85b,Wen90d} are closely related to each 
other~\cite{Wen95,Einarsson90,Oshikawa06,Sato06,Oshikawa07}.
Quantum spin liquids~\cite{Kalmeyer87,Levin05a,Kitaev06} and topological 
superconductors~\cite{Read00,Ivanov01,Kitaev01,Fu08,Qi11b,Schirmer22} are also 
promising 
platforms that harbor anyons, which are not only fundamentally interesting in 
their own right but also have attracted considerable attention for their
potential for quantum computation~\cite{Kitaev03,Nayak08}.

Understanding the behavior of quantum anyon gases
is a fundamentally
interesting problem. An exchange of $n$ Abelian anyons
with statistical angle $\theta$ gives the phase factor $e^{in^2\theta}$. This 
implies that $p$-tuples with $\theta=\pi(1-1/p)$ behave as bosons and thus may 
condense to form a superfluid~\cite{Laughlin88,Fetter89,Chen89,Wilczek90}. 
Historically, anyon superconductors, especially with $\theta=\pi/2$ 
(semions), have been extensively studied 
for their possible relevance to the high-$T_c$ 
superconductivity of the copper oxides~\cite{Halperin89,Kiefl90,Spielman90}. In
theoretical studies, the identification
of off-diagonal long-range order (ODLRO) is convincing evidence of 
superconductivity. The system of semions may be mapped to the 
$\nu=2$ integer quantum Hall system by trading the statistical fluxes for 
uniform magnetic ones. This mean field theory produces
algebraic ODLRO in the $2$-body reduced density
matrix~\cite{Girvin87,Jain89a}. There also exists an exactly solvable model of 
spinful semions beyond the mean field description,
which exhibits (not just algebraic) ODLRO~\cite{Girvin90,Girvin92}. In this 
paper, we revisit this problem by developing a method to construct the Hubbard 
model of anyons. 

The main goal of this work is to numerically confirm 
ODLRO 
for attractively interacting spinful semions 
in a lattice system. 
When fermions attractively interact, BCS pairs develop which are appropriately described by Bose-Einstein condensation (BEC) in the strong coupling limit~\cite{Leggett80,Nozieres85}.
Correspondingly, their ground states are 
expected to exhibit ODLRO in the two-body reduced density matrix. By comparing 
the behavior of ODLRO for semions and fermions, we investigate whether the BCS-BEC 
crossover occurs in semionic systems.
In the limit of large interaction strength, the system of 
semions 
is expected to exhibit the same behavior as a system of fermions with large Hubbard interaction strength, since pairs of semions and pairs of fermions both obey mutual Bose statistics.
In the weakly interacting regime, systems of fermions behave differently
from the BEC state and there is no clear connection between semionic
and fermionic systems.

We start with a construction of the Hubbard model of anyons
with tunable on-site interactions. This is not
a straightforward generalization of the tight-binding model for spinless 
anyons~\cite{Wen90c,Hatsugai91b,Hatsugai91,Kallin93,Kudo20,Kudo21,Kudo22} into
a spinful problem. Since anyons carry fractional statistics, 
two particles may not coincide even if their spins are different,
which makes it nontrivial to set up a model with a finite on-site Hubbard 
$U$ interaction. We overcome this difficulty by virtually splitting sites for 
each spin, thereby fixing the way that particles with different spin pass one another. 
Within this 
framework, we present a careful construction of the Hubbard model of anyons and
define 
the reduced density matrix for two semions. We numerically demonstrate clear 
ODLRO for semions in the strong coupling limit. 
We find that the 
reduced density matrices for semions and for fermions behave similarly in a 
wide range of $U$. This suggests that a BCS-BEC crossover
occurs in semionic systems.

\section{Hubbard model for anyons}
\subsection{Spinless anyons on a torus: the string gauge}
We begin by reviewing the hopping Hamiltonian of spinless anyons on a 
torus~\cite{Wen90c,Hatsugai91b,Hatsugai91,Kudo20}.
The statistical angle is set as $\theta/\pi=n/m$ with $n,m$ coprime. The 
Hilbert space for the system with $N$ particles is spanned by the basis 
$\ket{\{\bm{r}_k\};w}$, where $\{\bm{r}_k\}$ 
labels $N$ occupied sites and $w$ labels an additional internal degree of 
freedom required by topology of the torus. The label $w$ takes integer values 
from
1 to $m$. By modeling anyons as bosons with statistical fluxes, the Hamiltonian
with nearest-neighbor hopping is given 
by~\cite{Wen90c,Hatsugai91b,Hatsugai91,Kudo20}
\begin{align}
 H=-t\sum_{\langle ij\rangle}c_i^\dagger e^{i\theta_{ij}}W_{ij}c_j,
 \label{eq:ham_spinless}
\end{align}
where $c^\dagger_i$ is the creation operator for a hard-core boson on site $i$.
The hard-core condition is necessary to ensure consistency with the braid 
group: if particles can coincide, the system allows only Bose statistics~\cite{Wu84b}. The phase factor $e^{i\theta_{ij}}$ 
describes the statistical phase for a particle exchange. $W_{ij}$ is an
$m$-dimensional matrix associated with the degree of freedom $w$, which 
describes phase factors arising from global moves of anyons on a torus.
For $\theta/\pi=1$, $e^{i\theta_{ij}}$ produces the sign $\pm1$, 
resulting in $H$ 
reducing to the standard Hamiltonian of noninteracting fermions.

\begin{figure}[t!]
 \begin{center}
  \includegraphics[width=\columnwidth]{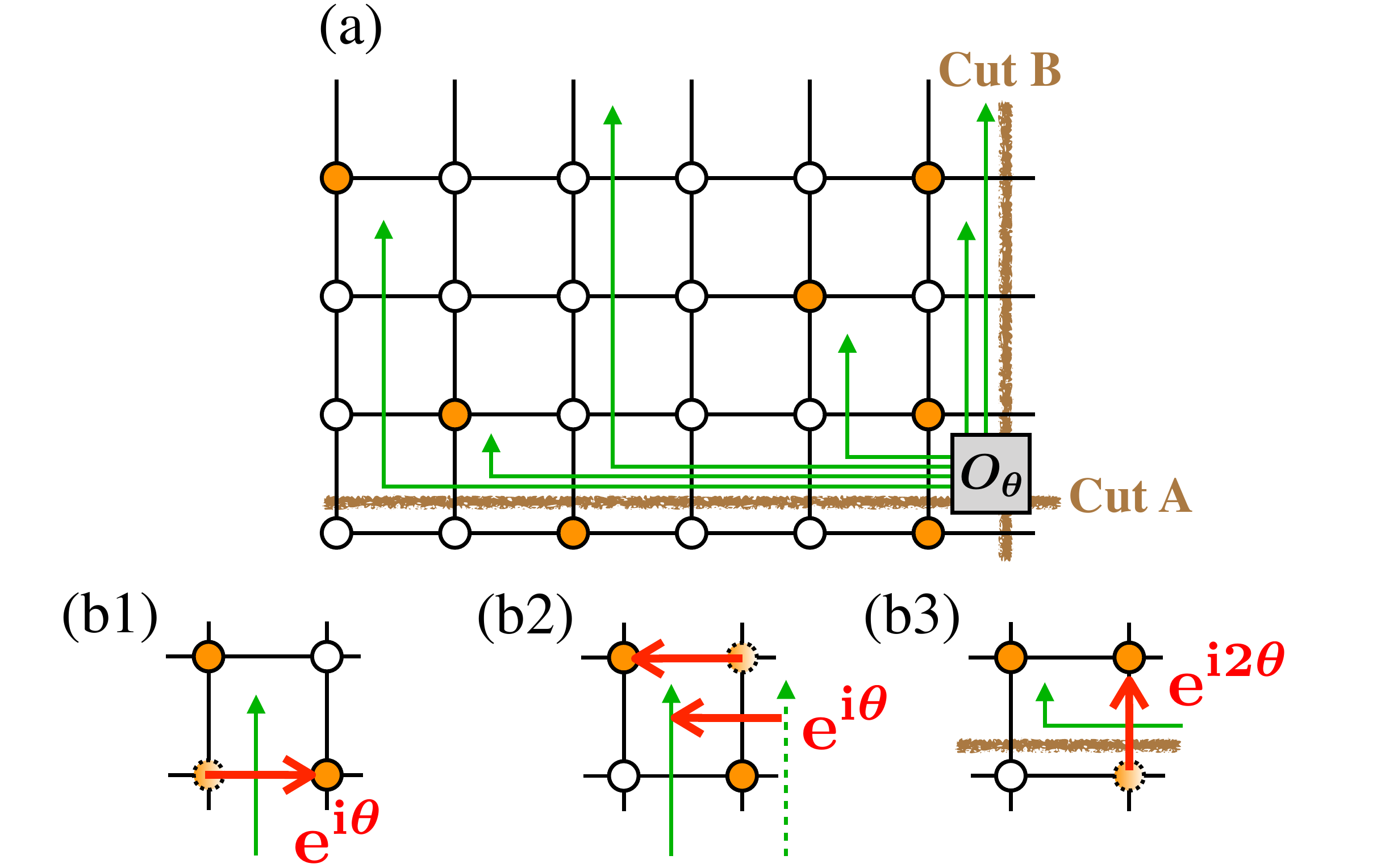}
 \end{center}
 \caption{(a) Sketch of a $6\times4$ lattice with $N=7$ particles with the 
 string gauge (green arrows). The brown lines represent the cuts A and B.
 (b1) Rule~$\theta_{ij}$:\eqref{item:local1}.
 (b2) Rule~$\theta_{ij}$:\eqref{item:local2}.
 (b3) Rule~$\theta_{ij}$:\eqref{item:twist_y}.
 }
 \label{fig:strings}
\end{figure}
The assignment of $\theta_{ij}$ is obtained by strings 
placed on the lattice as 
shown in Fig.~\ref{fig:strings}(a). These strings emanate from the origin 
represented by $O_\theta$, and terminate in the lower-right plaquette adjacent 
to each particle. They run first from right to left and then turn 
in the vertical direction 
at a 
proper plaquette. We also set cuts labeled A and B as shown in 
Fig.~\ref{fig:strings}(a). Using these strings, we first assign $\theta_{ij}$ 
and $W_{ij}$ for a hopping that does not cross the cuts:
\begin{enumerate}[$\theta_{ij}$:(i)]
 \setlength{\parskip}{0cm}
 \setlength{\itemsep}{0cm}
 \item If an anyon hops from left to right across a string, 
       it acquires the phase factor $e^{i\theta}$ 
       [Fig.~\ref{fig:strings}(b1)].
       \label{item:local1}
 \item If a string sweeps another anyon in the process of hopping, 
       the phase factor is gained
       as if the anyon crosses the string
       [Fig.~\ref{fig:strings}(b2)].
       \label{item:local2}
\end{enumerate}
\begin{enumerate}[$W_{ij}$:(i)]
 \setlength{\parskip}{0cm}
 \setlength{\itemsep}{0cm}
 \item The index $w$ does not change. This corresponds to 
       $W_{ij}=\bm{1}_m$ ($m$-dimensional identity matrix).
       \label{item:Wij}
\end{enumerate}
These rules ensure that the basis states $\ket{\{\bm{r}_k\};w}$ acquire the
statistical phase $e^{i\theta}$ whenever one takes two anyons and 
swaps their positions without crossing the cuts. 
For a hopping across the cuts, we need to set other rules as follows:
(Here we assume that an anyon hops across the cut A upward or the cut B from 
left to right.)
\begin{enumerate}[$\theta_{ij}$:(i)]
 \setcounter{enumi}{2}
 \setlength{\parskip}{0cm}
 \setlength{\itemsep}{0cm}
 \item (Cut A) The phase factor $e^{i2\theta}$ (not $e^{i\theta}$) is gained if
       an anyon crosses a string [Fig.~\ref{fig:strings}(b3)].
       \label{item:twist_y}
 \item (Cut A) $e^{iX\theta}$ is given, where $X$ is the number of other anyons
       that have the same $x$ coordinate as the hopping anyon.
       \label{item:order}
 \item (Cut B) One determines the phase factor obeying the 
       rules~\eqref{item:local1} and \eqref{item:local2} and then rearranges 
       the horizontal part of the string.
       \label{item:ex_x}
 \item (Cut B) The phase factor $e^{i(N-1)\theta}$ occurs.
       \label{item:twist_x}
\end{enumerate}
\begin{enumerate}[$W_{ij}$:(i)]
 \setcounter{enumi}{1}
 \setlength{\parskip}{0cm}
 \setlength{\itemsep}{0cm}
 \item (Cut A) The phase factor $e^{iw\theta}$ is given ($w$ is the additional 
       internal degree of freedom mentioned above). This corresponds to
       \begin{align}
	W_{ij}=W_y\equiv
	\text{diag}[e^{i2\theta},e^{i4\theta},\ldots,e^{i2m\theta}].
	\label{eq:Wy}
       \end{align}
       \label{item:global_y}
 \item (Cut B) The index $w$ is changed to $w-1$. This corresponds to 
       \begin{align}
	W_{ij}=W_x&\equiv\left[
	      \begin{array}{cccc} 
	       0 & 1 & \ldots & 0 \\
	       \vdots & \vdots & \ddots & \vdots \\
	       0 & 0 & \ldots & 1 \\
	       1 & 0 & \ldots & 0
	      \end{array}
	\right].
	\label{eq:Wx}
       \end{align}
       \label{item:global_x}
\end{enumerate}

The above rules for $\theta_{ij}$ imply that $\theta_{ij}$ can be written in
the following form:
\begin{align}
 \theta_{ij}=\sum_{k\neq i,j}A_{ij}^{k}n_k,
 \label{eq:thetaij_spinless}
\end{align}
where $A_{ij}^{k}$ is a real number determined from the rules given above and $n_k=c_k^\dagger c_k$.
The rules~$W_{ij}$:\eqref{item:global_y} and \eqref{item:global_x} are required
for some algebraic constraints of the braid group on a torus: operators 
$\tau_i$ and $\rho_i$ that move the $i$th particle along a 
noncontractible loop in the $x$ and $y$ direction, respectively, satisfy 
$\rho_i^{-1}\tau_j\rho_i\tau_j^{-1}=B_{ij}$, where $B_{ij}$ is an operator that
moves the $i$th particle around the $j$th particle along a closed loop. This is
consistent with the fact that we have $B_{ij}=e^{i2\theta}$ and
\begin{align}
 W_y^{-1}W_xW_yW_x^{-1}=e^{i2\theta}.
 \label{eq:WWWW}
\end{align}
The rules~$\theta_{ij}$:\eqref{item:twist_y} and \eqref{item:twist_x} also 
play a role to remove the artificial twisted boundary condition caused by the 
anyon fluxes. The twisted boundary conditions are introduced by modifying the
matrices as $W_x\rightarrow e^{i\eta_x}W_x$ and 
$W_y\rightarrow e^{i\eta_y}W_y$, where $\eta_x$ and $\eta_y$ are the 
twisted boundary condition angles. 

\subsection{Generalization to spinful system}
Let us generalize the above formulation into a system of spinful anyons
on a lattice. The statistical phase for an exchange of anyons with the 
same spin is set as $e^{i\theta}$. Note that the statistical
phase is not well defined for an exchange of opposite spins since this 
operation does not form a closed loop in configuration space.
Instead, two
consecutive operations, which is equivalent to a move of a spin 
$\uparrow$ around a spin $\downarrow$ or vice versa (this is referred to as 
``local move'' below), forms a closed loop in the
configuration space. We assign to this the phase factor $e^{i2\theta}$, i.e.,
a local move of a particle around another particle produces the phase 
factor $e^{i2\theta}$ whether or not they have the same spin.

Incorporating the on-site interaction between anyons with opposite spins, we 
write the Hamiltonian in the form of the Hubbard model:
\begin{align}
 H=-t\sum_{\langle ij\rangle}\sum_{\alpha=\uparrow,\downarrow}
 c_{i\alpha}^\dagger e^{i\theta_{ij\alpha}}W_{ij\alpha}c_{j\alpha}
 +U\sum_in_{i\uparrow}n_{i\downarrow},
 \label{eq:ham_spinful}
\end{align}
where $n_{i\alpha}=c_{i\alpha}^\dagger c_{i\alpha}$ and $c_{i\alpha}^\dagger$ 
is the creation operator for a spinful hard-core boson
satisfying $c_{i\uparrow}^{\dagger2}=c_{i\downarrow}^{\dagger2}=0$. 
As in the spinless case, $\theta_{ij\alpha}$ depends on the number 
operators of all sites and, therefore, $e^{i\theta_{ij\alpha}}$ generates 
complicated non-local many-body interactions. This nature makes it quite
difficult to apply effective formalisms such as the Bogoliubov-de Gennes
treatment.
The construction of $\theta_{ij\alpha}$ and $W_{ij\alpha}$ is described below.

\subsubsection{$U=+\infty$}
The system with the hard-core constraint 
$c_{i\uparrow}^\dagger c_{i\downarrow}^\dagger=0$ (equivalently $U=+\infty$)
is given by a simple generalization of the formulation 
for spinless anyons~\cite{Kudo22}. The 
Hilbert space is spanned by the basis $\ket{\{\bm{r}_j\},\{\bm{r}_k\};w}$ with 
$j=1,\ldots,N_\uparrow$ and 
$k=1,\ldots,N_\downarrow$, where $N_\alpha$ is the particle number with spin 
$\alpha$, $\{\bm{r}_j\}$ and $\{\bm{r}_k\}$ label
sites occupied by spin $\uparrow$ and $\downarrow$, respectively. Here the sets of positions
$\{\bm{r}_j\}$ and $\{\bm{r}_k\}$ are
disjoint. One can construct this from the spinless basis $\ket{\{\bm{r}_l\};w}$
with $l=1,\ldots,N$ by partitioning positions into two groups of $N_\uparrow$ 
and $N_\downarrow$ each, where $N=N_\uparrow+N_\downarrow$. Due to the multiple
ways of making this partition, the dimension of the
Hilbert space is $N!/(N_\uparrow!N_\downarrow!)$ times larger than that for
the corresponding spinless system. Using this basis, we consider the 
Hamiltonian in Eq.~\eqref{eq:ham_spinful} with
\begin{align}
 &\theta_{ij\uparrow}=\theta_{ij\downarrow}=\theta_{ij},
 \label{eq:thetaijalpha_NoDO}\\
 &W_{ij\uparrow}=W_{ij\downarrow}=W_{ij}
 \label{eq:Wijalpha_NoDO}
\end{align}
where $\theta_{ij}$ and $W_{ij}$ have been defined in the spinless problem 
(we redefine $\theta_{ij}$ in Eq.~\eqref{eq:thetaij_spinless} with 
$n_k=n_{k\uparrow}+n_{k\downarrow}$). As
ensured by the above rules~\eqref{item:local1}-\eqref{item:global_x}, the phase
factors $e^{i\theta}$ and $e^{i2\theta}$ are given for 
an exchange of particles with the same spin and for a local move of a particle around another, 
respectively. This system has SU(2) spin-rotational symmetry since 
$\theta_{ij\alpha}$ 
is invariant under the transformation
$\bm{c}_i^\dagger\rightarrow\bm{c}_i^\dagger u$ with $u\in\text{SU(2)}$, where
$\bm{c}_i^\dagger\equiv(c_{i\uparrow}^\dagger,c_{i\downarrow}^\dagger)$.

\subsubsection{Finite $U$}
\begin{figure}[t!]
 \begin{center}
  \includegraphics[width=\columnwidth]{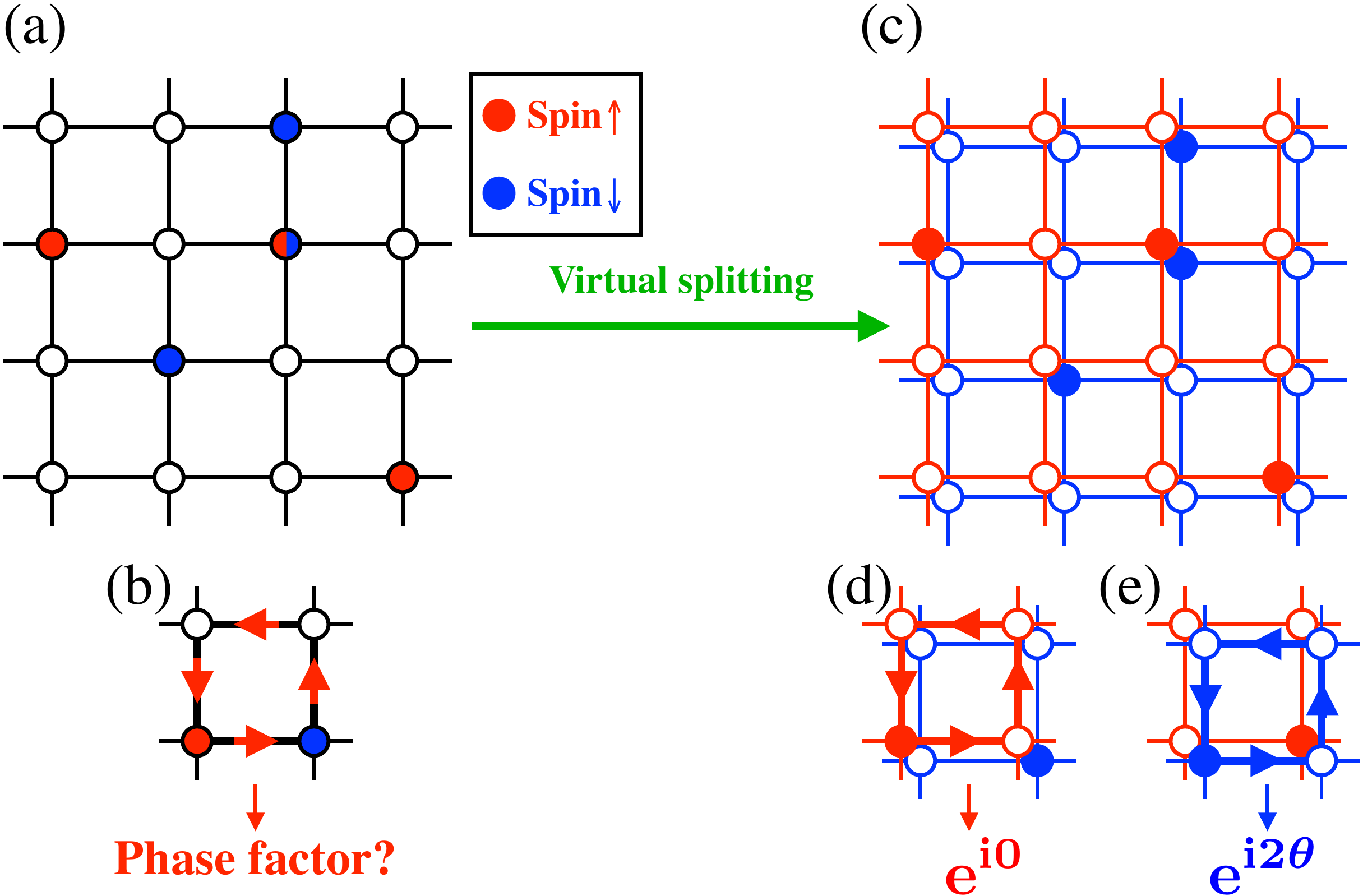}
 \end{center}
 \caption{(a) Sketch of a $4\times4$ lattice with the particle numbers
 $N_\uparrow=N_\downarrow=3$. (b) Local move of an anyon with spin $\uparrow$.
 The red arrows represent the direction of the hopping.
 The phase factor is generally not well-defined due to the coincidence of spin-up and spin-down particles along the path of the spin-up particle. (c) Virtual 
 splitting. (d)-(e) Local moves of anyons with (d) spin $\uparrow$ and (e) spin 
 $\downarrow$. The bold lines and its arrows represent the paths and the 
 direction of the hoppings. Because of the virtual splitting, each phase factor
 is uniquely determined.
 }
 \label{fig:split}
\end{figure}

The system with finite $U$ allows for double occupancy of sites
as shown in Fig.~\ref{fig:split}(a). We note here that whether or not such a system 
can be well-defined is itself a nontrivial problem. As
mentioned above, moving a spin $\uparrow$ around a spin $\downarrow$ or vice 
versa along a closed loop gives the phase factor $e^{i2\theta}$, 
implying that the 
positions of particles are singular points unless $e^{i2\theta}=1$.
This feature 
appears to prohibit an unambiguous identification of the phase factor for a 
local move shown in Fig.~\ref{fig:split}(b), where two anyons with opposite 
spins coincide in the process.

This ambiguity can be resolved by virtually splitting the sites and thereby 
fixing the way the particles
pass around each other. As shown in Fig.~\ref{fig:split}(c), let us 
shift the sites for spin $\downarrow$ 
particles infinitesimally in the south east direction.
This configuration enables
us to see whether anyons are enclosed or not for a given local move. 
Accordingly, the phase factor is uniquely 
determined since the path of the spin $\uparrow$ particle does not enclose the spin $\downarrow$ particle in Fig.~\ref{fig:split}(d), whereas the path of the spin $\downarrow$ particle encloses the spin $\uparrow$ particle, as seen in Fig.~\ref{fig:split}(e). 

\begin{figure}[t!]
 \begin{center}
  \includegraphics[width=\columnwidth]{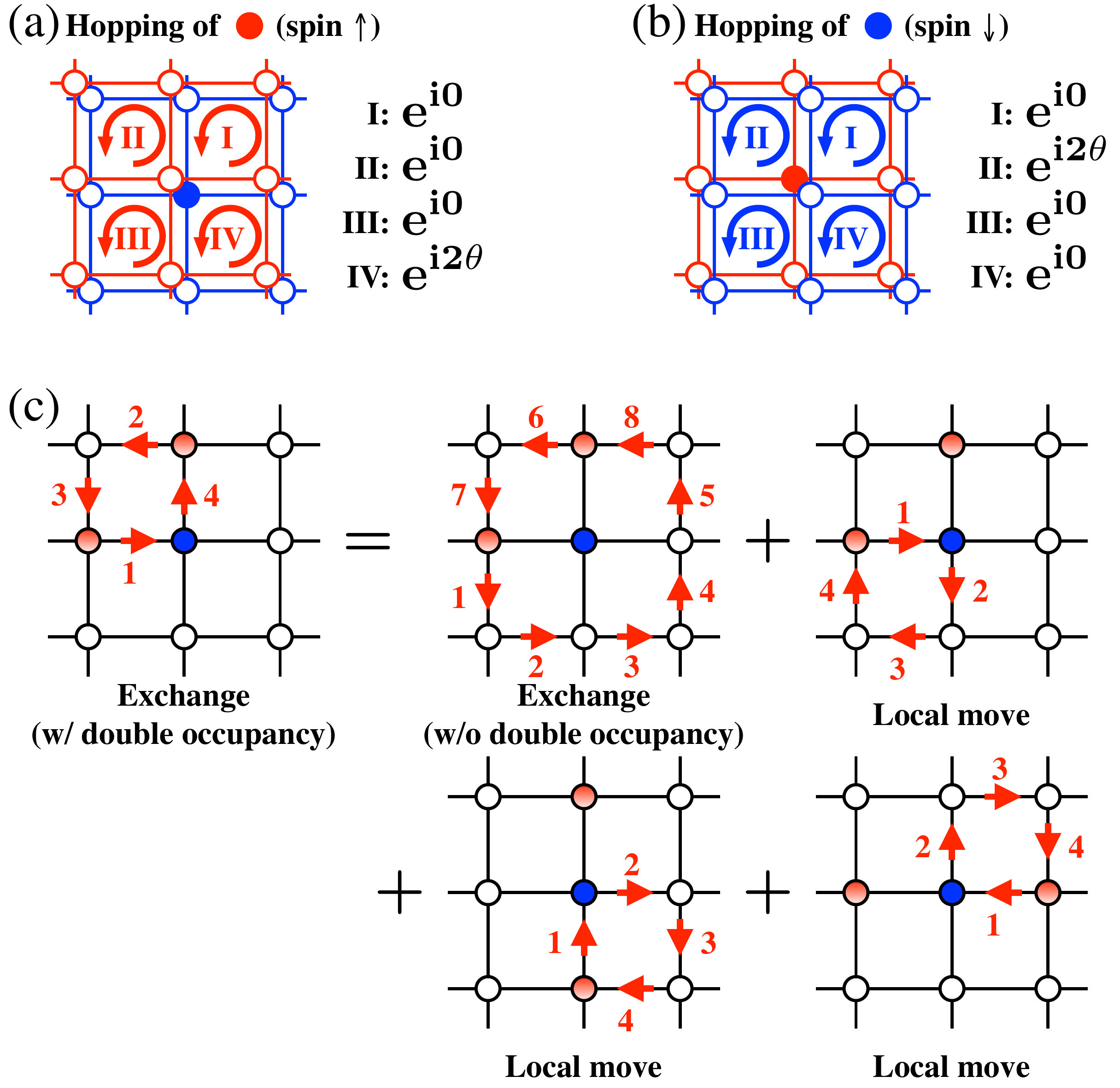}
 \end{center}
 \caption{
(a)(b) Four plaquettes adjacent to an anyon with (a) spin $\downarrow$ and (b)
 spin $\uparrow$. Each plaquette is labeled by I$,\ldots,$IV. The hoppings of 
 (a) spin $\uparrow$ and (b) spin $\downarrow$ along each 
 plaquette give the phase factors as listed on the right in each figure.
 (c) An exchange operation of two anyons with spin $\uparrow$ is decomposed 
 into four
 parts: an exchange with no double occupancy on any site and local 
 moves involving the double occupancy. The numbers written next to the arrows
 indicate the order of operation.
 }
 \label{fig:plaquette}
\end{figure}
We have not included terms with $n_{i\alpha}$ or 
$n_{j\alpha}$ in $\theta_{ij\beta}$ in Eq.~\eqref{eq:thetaijalpha_NoDO} so far 
because of the hard-core nature of the anyons.
To incorporate the virtual splitting, we now add terms such as 
$\theta_{ij\alpha} =\theta_{ij}+D_{ij\alpha}$, where
\begin{align}
 &D_{ij\uparrow}
 =
 A_{ij}^{i\downarrow}n_{i\downarrow}
 +A_{ij}^{j\downarrow}n_{j\downarrow}
 \label{eq:thetaiju_DO},\\
 &D_{ij\downarrow}
=
 A_{ij}^{i\uparrow}n_{i\uparrow}
 +A_{ij}^{j\uparrow}n_{j\uparrow}
 \label{eq:thetaijd_DO}.
\end{align}
The coefficients $A_{ij}^{i\alpha}$ are assigned as follows. We 
put an anyon 
with spin $\downarrow$ on a site, see Fig.~\ref{fig:plaquette}(a). We then 
design rules of 
$A_{ij}^{i\downarrow}$ so that the phase accumulated by moving spin 
$\uparrow$ along each adjacent plaquette matches with the virtual splitting.
Each phase is listed on the right in the figure.
We also determine 
$A_{ij}^{i\uparrow}$ in the same way, see Fig.~\ref{fig:plaquette}(b). 
These conditions regarding only four adjacent plaquettes are sufficient to 
design $A_{ij}^{i\alpha}$ as can be seen from the following example. An 
exchange operation shown in Fig.~\ref{fig:plaquette}(c), where double 
occupancy occurs, can be decomposed into several operations that are well-defined 
in the above framework: an exchange operation involving no double 
occupancy and local moves along a plaquette. This implies that the exchange
operation before the decomposition should give the proper phase that reflects 
the virtual splitting. Other operations in which double occupancy occurs 
also give proper phase factors since these can be decomposed in the same way as
above. 

\begin{figure}[t!]
 \begin{center}
  \includegraphics[width=\columnwidth]{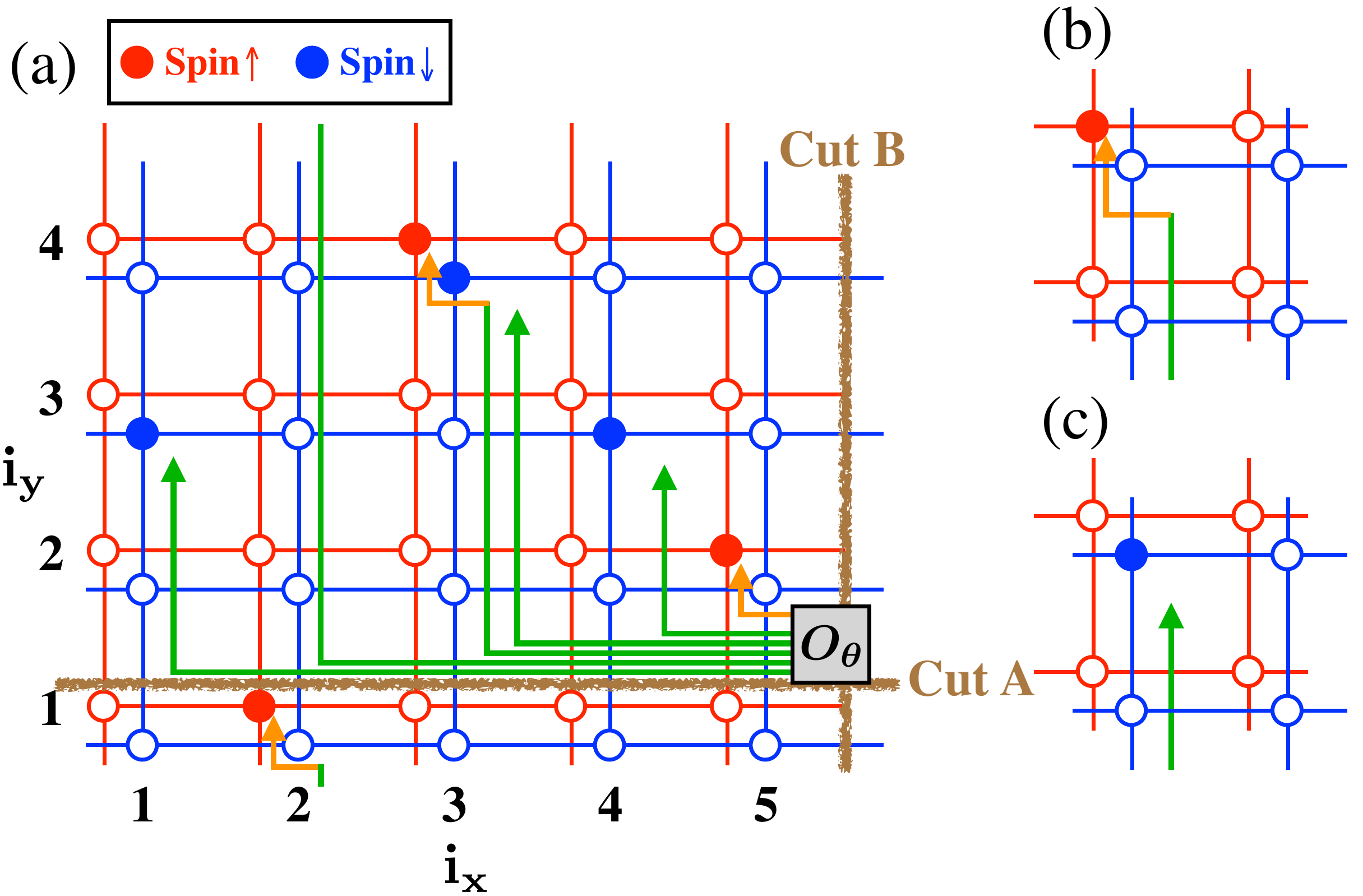}
 \end{center}
 \caption{
 (a) Sketch of a $5\times4$ lattice with $N_\uparrow=N_\downarrow=3$ with the
 string gauge (green or orange arrows). The brown lines represent the cut A and
 the cut B.
 (b)(c) String for (b) spin $\uparrow$ and (c) spin $\downarrow$.
 }
 \label{fig:plaquette_vir}
\end{figure}
We now give an explicit construction of $A_{ij}^{i\alpha}$ by modifying the
rules for the strings. Figure~\ref{fig:plaquette_vir}(a) pictorially shows the
strings attached to each particle with the virtual splitting. 
The assignment of strings is almost the same as the spinless case, since we do not
change the form of $\theta_{ij\alpha}$ in Eq.~\eqref{eq:thetaijalpha_NoDO}
except for the addition of new terms $D_{ij\alpha}$. The difference is in the 
vicinity of particles as shown in Figs.~\ref{fig:plaquette_vir}(b) and (c), 
where we introduce a new orange string around spin $\uparrow$. Using these, we 
employ the previous 
rules~$\theta_{ij}$:\eqref{item:local1}-\eqref{item:twist_x}, but
change only the rule~$\theta_{ij}$:\eqref{item:twist_y} as 
\begin{enumerate}[$\theta_{ij}$:(i')]
 \setcounter{enumi}{2}
 \setlength{\parskip}{0cm}
 \setlength{\itemsep}{0cm}
 \item (Cut A) The phase factor $e^{i2\theta} (e^{i\theta})$ is given if 
       an anyon crosses a green (orange) string.
       \label{item:twist_y_spinful}
\end{enumerate}
We still use Eq.~\eqref{eq:Wijalpha_NoDO} for $W_{ij\alpha}$ without any
modification. 

\begin{figure}[t!]
 \begin{center}
  \includegraphics[width=\columnwidth]{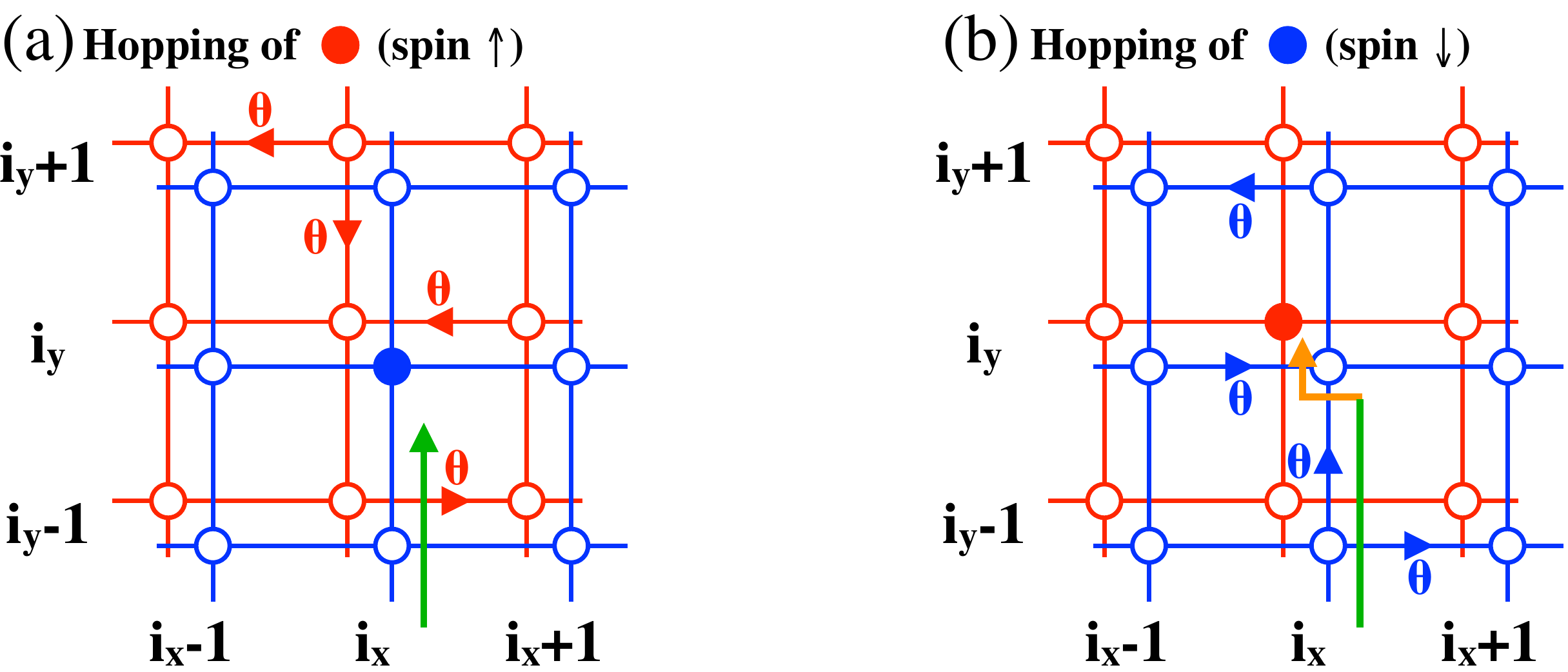}
 \end{center}
 \caption{
 Hopping phases of an anyon with (a) spin $\uparrow$ and (b) spin
 $\downarrow$. The label of the site $(i_x,i_y)$ satisfies $3\leq i_y\leq N_y$ 
 (no restriction on $i_x$). The sum of phases along each link is consistent
 with the virtual splitting.
 }
 \label{fig:check1}
\end{figure}
The above rules clearly produce the phase factor shown in 
Figs.~\ref{fig:plaquette}(a) and (b). Figure~\ref{fig:check1} shows the hopping
phases assigned to each link.
Here we assume $3\leq i_y\leq N_y$ (no restriction on $i_x$), where
$i_{x(y)}$ is the label of sites, see 
Fig.~\ref{fig:plaquette_vir}(a). Clearly, the sum of the hopping phases along
each plaquette matches with the virtual splitting. Other 
situations (namely, $\vec{i}$ with any $i_x$ and $1\leq i_y\leq2$) are 
discussed in
Appendix~\ref{appx:check}.

Because of the virtual splitting, $\theta_{ij\alpha}$ no longer takes the form
$\sum_{k\neq i,j}A_{ij}^{k}(n_{k\uparrow}+n_{k\downarrow})$. This breaks SU(2) 
spin-rotational symmetry unless $\theta/\pi$ is an integer. 
If particles are
fermions or bosons, our model reduces to the standard Hubbard model or the 
spinful Bose-Hubbard model with hard-core interactions between bosons with the 
same spin.
Although the absence of SU(2) symmetry for anyons  is just an artificial
effect due to the virtual splitting, we stress that the fractional statistics 
is well-defined without ambiguity in our model. This method enables us to
effectively investigate an interplay between the short-range interaction and 
the fractional statistics for systems that we can deal with using the exact 
diagonalization method.

\subsubsection{$U\ll-1$}
Let us discuss the effective Hamiltonian in the strong interaction limit 
$U\ll-1$. We denote the 
first and second terms in 
Eq.~\eqref{eq:ham_spinful} by $H_\text{kin}$ and $H_\text{int}$, respectively, 
and we consider $H_\text{kin}$ as a perturbation to $H_\text{int}$. In the 
unperturbed problem,
the ground state forms pairs of anyons at each site. This 
macroscopic degeneracy is lifted by the second-order perturbation: 
$H'=PH_\text{kin}Q\frac{1}{E_0-H_\text{int}}QH_\text{kin}P$, where $P$ is the 
projection operator to the degenerate ground state, $H_\text{int}P=E_0P$ and 
$Q=1-P$. 

For $\theta/\pi=1$ (fermions), we have 
$H'=(2t^2/U)\sum_{\langle ij\rangle}b^\dagger_ib_j$ as shown in 
Appendix~\ref{appx:Hfermions}, where $b_i^\dagger$ 
is the creation operator for a hard-core boson. We ignore a term which is 
constant when the particle number is fixed. The
boundary conditions of $H'$ and $H$ satisfy 
$(\eta'_x,\eta_y')=(2\eta_x,2\eta_y)$.
The expression of $H'$ for general $\theta$ is complicated but we can 
deduce it by considering the statistics of a pair. For instance, pairs of 
semions with $\theta/\pi=1/2$ behave like bosons, implying that the 
effective Hamiltonian is given by
\begin{align}
 H'=(2t^2/U)\sum_{\langle ij\rangle}b^\dagger_ib_j\bm{1}_2,
 \label{eq:H'boson}
\end{align}
where the identity matrix $\bm{1}_2$ comes from $W_x^2=W_y^2=\bm{1}_2$ [see
Eqs.~\eqref{eq:Wy} and \eqref{eq:Wx} with $m=2$], which leads to a
2-fold degeneracy. The boundary conditions satisfy 
$(\eta'_x,\eta_y')=(2\eta_x+\pi,2\eta_y+\pi)$. The shift of $\pi$ follows from
the fact that the rules we 
constructed using strings produce the 
phase factor $-1$ apart from the twisted boundary conditions
when one moves a pair of semions along a noncontractible loop
in the $x$ or $y$ direction~\cite{com}.

\subsection{Two-body reduced density matrix}
To investigate the presence of ODLRO for these systems,
we now define the two-body reduced density matrix as ~\cite{Jain89a,Girvin92}
\begin{align}
 &\rho(i,j)
 =\bra{\Psi}
 T_{ij}
 \ket{\Psi},\\
 &T_{ij}=\prod_{\langle kl\rangle\in L_{ij}}
 \left[
 c_{k\uparrow}^\dagger 
 e^{i\theta_{kl\uparrow}}W_{kl\uparrow}
 c_{l\uparrow}
 c_{k\downarrow}^\dagger e^{i\theta_{kl\downarrow}}W_{kl\downarrow}
 c_{l\downarrow}
 \right]
\end{align}
where $\ket{\Psi}$ is the ground state of $H$ and
$\prod_{\langle kl\rangle\in L_{ij}}$ indicates the products over all the links
of a path, $L_{ij}$, starting from $j$ to $i$. For a given basis state
$\ket{\{\bm{r}_j\},\{\bm{r}_k\};w}$,
we choose the path $L_{ij}$ so that (i) there is no particle along it except for
the site $j$
and (ii)
all paths for any basis states are continuously connected with each other
(i.e., two paths for two basis states always form a contractible loop on a
torus).

The value of $\rho(i,j)$ is independent of the choice of the path $L_{ij}$
if $\theta/\pi=\mathbb{Z}/2$. To see 
this, consider two paths $L_{ij}$ and $L_{ij}'$. Since the two paths form a
closed loop, denoted by $(L_{ij}')^{-1}L_{ij}$, the corresponding operators
$T_{ij}$ and $T_{ij}'$ satisfy
\begin{align}
 T_{ij}'^{-1}T_{ij}\ket{\{\bm{r}_k\},\{\bm{r}_l\};w}
 =e^{i4n\theta}\ket{\{\bm{r}_k\},\{\bm{r}_l\};w},
\end{align}
where $n$ is the number of anyons enclosed by the closed loop, and the sites 
$i$ and $j$ are assumed to be occupied in the basis state.
This implies
$T_{ij}\ket{\{\bm{r}_k\},\{\bm{r}_l\};w}
=T'_{ij}\ket{\{\bm{r}_k\},\{\bm{r}_l\};w}$ for $\theta/\pi=\mathbb{Z}/2$, i.e.,
$\rho(i,j)$ is path-independent.
[For $\theta/\pi=1$, 
$\rho(i,j)$ is equivalent to the standard reduced matrix for fermions.]

Anyonic systems should be invariant under a many-particle translation since the
statistical gauge field should depend only on relative coordinates of each 
pair of particles~\cite{Wilczek82}. This implies that our Hamiltonian satisfies
$GTH(\eta_x,\eta_y)T^\dagger G^\dagger=H(\eta_x,\eta_y)$, where $T$ is a 
translational operator and $G$ is a gauge transformation that rearranges the
configuration of strings. 
Since $|\rho(i,j)|$ is a gauge invariant, the 
relation $|\rho(i,j)|=|\rho(i+\delta,j+\delta)|$ should hold for any 
translation $\delta=(\delta_x,\delta_y)$. We numerically confirm this as 
mentioned below.

\section{Results}
We consider a system with 6 fermions or 6 semions on a $6\times6$ lattice
by using the above setup. 
The inter-particle distance is $r_0=a\sqrt{(6\times6)/6}\approx2.45a$, 
where $a$ is the lattice constant. We set $t=1$ and $U\leq0$. 
The system of attractively interacting fermions produces superconducting 
states, which
is demonstrated by the appearance of ODLRO (discussed below). We also compute 
the superconducting order parameter in Appendix~\ref{appx:BCS}.
By comparing the 
reduced density matrix $\rho(i,j)$ for semions and fermions, we identify the 
emergence of semion superconductivity.

\subsection{Spectral flow}
\begin{figure}[t!]
 \begin{center}
  \includegraphics[width=\columnwidth]{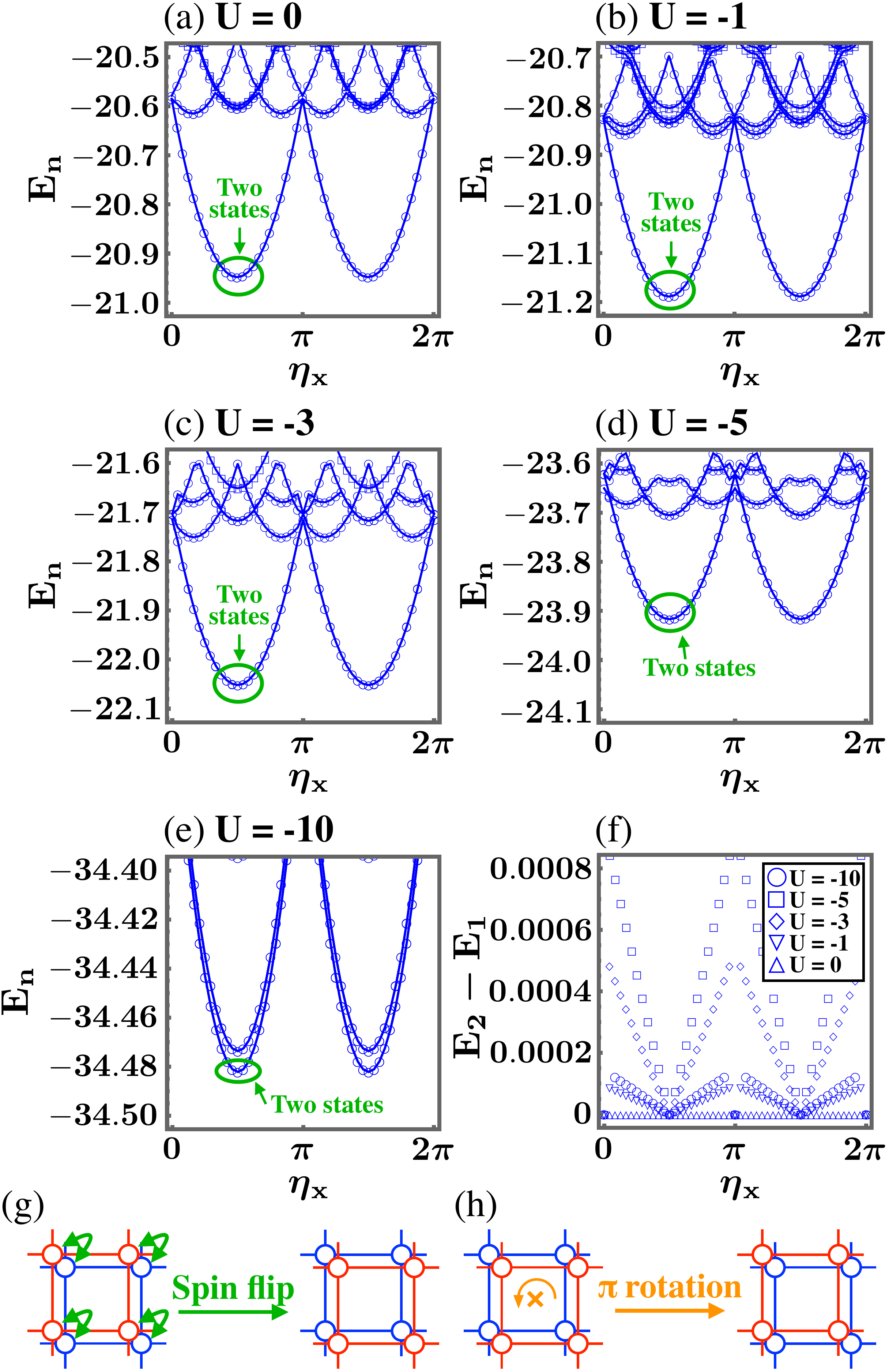}
 \end{center}
 \caption{
 (a)-(e) Spectral flows for semions at (a) $U=0$, (b) $U=-1$, (c) $U=-3$, (d) $U=-5$ and 
 (e) $U=-10$. The system size is $6\times6$ and the particle number is $N=6$. 
 The twist angle in the $y$-direction is set as $\eta_y=0$. The circles indicate 
 $S_z=0$. The squares that appear only in (a)-(c) indicate $S_z=1$. 
 States with larger $S_z$ not shown here have higher energies outside the 
 figure. The lowest 5 energies at each $\eta_x$ and $S_z$ are plotted. 
 (f) Energy gap between the two states indicated by the green circles in 
 (a)-(e). 
 (g)(h) Pictorial explanation of the unitary operator $R$: (g) the spin-flip 
 and (h) the $\pi$ rotation.
 }
 \label{fig:spectrum}
\end{figure}
We first investigate the energy spectrum for semions. 
In Figs.~\ref{fig:spectrum}(a)-(e), we plot the energy as a function of the 
twisted boundary condition angle in the $x$-direction $\eta_x$. We here set $\eta_y=0$. The ground 
state always gives $S_z=0$ at any $\eta_x$. The 
eigenenergies have a periodicity $E_n(\eta_x,\eta_y)=E_n(\eta_x+\pi,\eta_y)$ 
at any $U$. This follows from a gauge transformation 
$W_x=e^{i\pi}G_1W_xG_1^\dagger$~\cite{Hatsugai91}, where
$G_1=\text{diag}[1,-1]$. This leads to the relation 
$G_1H(\eta_x,\eta_y)G_1^\dagger=H(\eta_x+\pi,\eta_y)$. We also note 
that in Figs.~\ref{fig:spectrum}(a)-(e), there appears to be a unique 
low-energy state separated by a gap except for $\eta_x\approx0$ 
(equivalently $\eta_x\approx\pi$), but 
in fact there are two states as indicated by the green text. In 
Fig.~\ref{fig:spectrum}(f), we plot the energy gap between the two states. At 
$U=0$, 
the ground state is doubly degenerate at any $\eta_x$. For $U<0$, on the other
hand, the gap is very small but finite and it closes at 
$\eta_x=\pi/2$ (equivalently $3\pi/2$).

The gap closing at $\eta_x=\pi/2$ is explained by symmetry as follows. Our 
Hamiltonian satisfies 
$G_2RH(\eta_x,\eta_y)R^\dagger G_2^\dagger=H(-\eta_x,-\eta_y)$, where the 
operator $R$ switches the spin orientation [Fig.~\ref{fig:spectrum}(g)] and 
then rotates the system by $\pi$ [Fig.~\ref{fig:spectrum}(h)], and $G_2$ 
is a gauge transformation that rearranges strings. Combining it with the gauge 
transformation $G_1$ discussed above, one obtains 
$SH(\eta_x,\eta_y)S^\dagger=H(-\eta_x-\pi,-\eta_y)$ with $S\equiv G_2RG_1$, 
i.e., 
\begin{align}
 [H(\pi/2,0),S]=0.
\end{align}
In order to demonstrate the origin of the gap closing at $\eta_x=\pi/2$, we 
would like to identify the simultaneous eigenstate of $S$ as
$S\ket{\lambda_j}=e^{i\lambda_j}\ket{\lambda_j}$ ($j=1,2$) at $\eta_x=\pi/2$.
However, this is difficult to do since we do not know an explicit form of 
$G_2$. Instead, we now compare
the values of $\bra{\Psi}n_i\ket{\Psi}$ and 
$\bra{\Psi'}n_i\ket{\Psi'}$, where 
$\ket{\Psi}=\sum_{j=1}^2\psi_j\ket{\lambda_j}$ 
is a general state in the doubly degenerate state subspace and 
$\ket{\Psi'}\equiv RG_1\ket{\Psi}
=\sum_{j=1}^2e^{i\lambda_j}\psi_jG_2^\dagger\ket{\lambda_j}$. 
Noting that gauge transformations do not change the physical observables, i.e.,
$G_2n_iG_2^\dagger=n_i$, one 
obtains $\bra{\Psi'}n_i\ket{\Psi'}=\sum_{jk}e^{i(-\lambda_j+\lambda_k)}
\psi_j^*\psi_k\bra{\lambda_j}n_i\ket{\lambda_k}$. This implies that if 
$\lambda_1=\lambda_2$, we have 
$\bra{\Psi'}n_i\ket{\Psi'}=\bra{\Psi}n_i\ket{\Psi}$. We numerically confirm 
$\bra{\Psi'}n_i\ket{\Psi'}\neq\bra{\Psi}n_i\ket{\Psi}$ at $U=-1$ with the site 
$i=(i_x,i_y)=(1,1)$, which leads to $\lambda_1\neq\lambda_2$ by the 
contrapositive of the above statement. This demonstrates that the gap 
closing in Fig.~\ref{fig:spectrum}(f) is a result of the symmetry $S$.

We briefly mention other degeneracies. As shown in 
Figs.~\ref{fig:spectrum}(a)-(e), many states are degenerate at 
$\eta_x=0$. We expect that this is characterized by some combinations of 
operations such as $R$, a $\pi/2$ rotation, mirror transformation (gauge transformations would also be required to rearrange strings).
As suggested by Fig.~\ref{fig:spectrum}(e), we obtain a $4$-fold
(quasi)degeneracy for a sufficiently large interaction with $\eta_y=0$. This is
consistent with the energy spectrum of the effective 
Hamiltonian in Eq.~\eqref{eq:H'boson}, where the ground state is 2-fold 
degenerate apart from $\bm{1}_2$. The origin of 
the two-fold degeneracy shown in Fig.~\ref{fig:spectrum}(a) at arbitrary 
$\eta_x$ is still an open question.

\subsection{Two-body reduced density matrix}
\begin{figure}[t!]
 \begin{center}
  \includegraphics[width=\columnwidth]{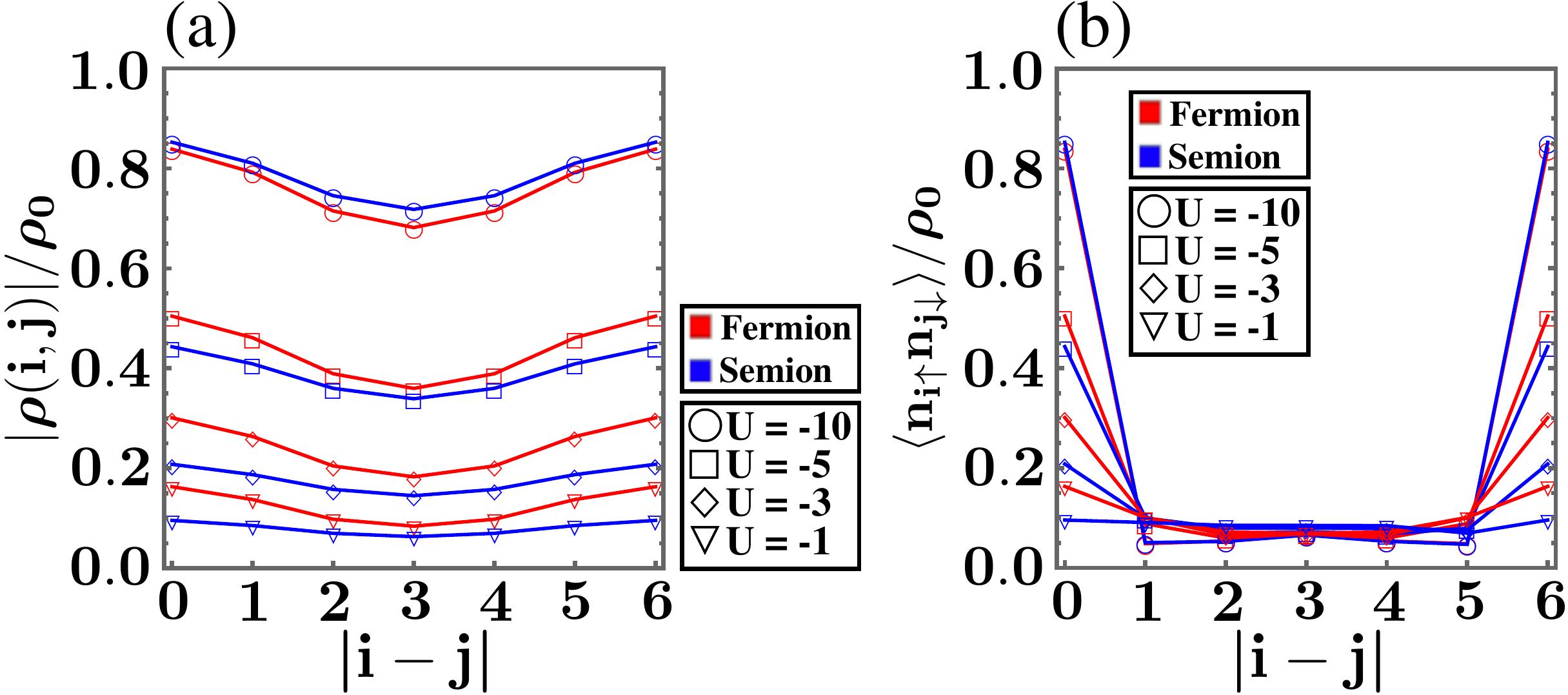}
 \end{center}
 \caption{
 (a) Absolute value of the two-body reduced density matrix $\rho(i,j)$ and (b) 
 the density-density
 correlation function $\langle n_{i\uparrow}n_{j\downarrow}\rangle$ as 
 functions of the distance $|i-j|$. Each are normalized by the density 
 $\rho_0\equiv N/(2N_xN_y)$. The system size is $N_x\times N_y=6\times6$ and 
 the particle number is $N=6$. We set $(\eta_x,\eta_y)=(\pi/4,0)$ and $(\eta_x,\eta_y)=(0,0)$
 for semions and fermions, respectively.
 }
 \label{fig:rhoij}
\end{figure}
Let us now discuss the presence of off-diagonal long-range order. Focusing on
the systems at $U<0$, we plot in Fig.~\ref{fig:rhoij}(a) the two-body reduced 
density matrix $|\rho(i,j)|/\rho_0$, where $\rho_0=N/(2N_xN_y)=1/12$ is the 
density. We set $i=(i_x,i_y)=(1,2)$ and change the other site as 
$j=(1,2),(2,2),(3,2),\ldots,(6,2)$, corresponding to $|j-i|=0,1,2,\ldots,6$. 
The twisted boundary conditions are set as $(\eta_x,\eta_y)=(\pi/4,0)$ to 
obtain a unique ground state. As a sanity check for translational 
symmetry, we confirm that the values of $|\rho(i,j)|$ at $U=-1$ with $i=(5,3)$ 
and $j=(5,3),(6,3),\ldots$ are in agreement with that in 
Fig.~\ref{fig:rhoij}(a).
In the figure, we also show the result with the same system but with fermions 
with $(\eta_x,\eta_y)=(0,0)$ as a reference state of the superconductor.

The size of the Cooper pair in the fermionic system, according to mean-field theory, coincides with the 
inter-particle distance at $U=-2.56$ as shown in Appendix~\ref{appx:BCS}. Based
on this, we show four types of data in Fig.~\ref{fig:rhoij}(a): the BEC
limit ($U=-10$), the BEC regime ($U=-5$), the intermediate regime ($U=-3$), and
the BCS regime ($U=-1$). In the BEC limit ($U=-10$), $\rho(i,j)$ for
semions and fermions has quantitatively similar 
structure, demonstrating off-diagonal long-range order of semionic superconductivity.
In Fig.~\ref{fig:rhoij}(b), we plot the density-density correlation
function $\langle n_{i\uparrow}n_{j\downarrow}\rangle$ in the same settings as 
Fig.~\ref{fig:rhoij}(a). The result for semions at $U=-10$ is almost the same 
as that for fermions and one can also see 
$\langle n_{i\uparrow}n_{i\downarrow}\rangle\sim\rho_0$ in both data.
These are consistent with the expectation that a pair of semions obeys 
Bose statistics. 
As the interaction becomes weaker, 
$\langle n_{i\uparrow}n_{i\downarrow}\rangle$ becomes smaller, implying that 
the size of the pairs is larger (since the density-density correlation function obeys the normalization condition $\sum_j \langle n_{i \uparrow}n_{j \downarrow} \rangle= N_\uparrow N_\downarrow /( N_x N_y)$ where $N_\uparrow$ and $N_\downarrow$ are the number of up-spin and down-spin particles, respectively). The calculation of
$\langle n_{i\uparrow}n_{j\downarrow}\rangle$ is
useful to estimate the size of pairs, although the system size that we can 
currently access is not large enough to do it quantitatively.

In the BEC ($U=-5$), the intermediate ($U=-3$), and the BCS ($U=-1$) regimes,
$\rho(i,j)$ and $\langle n_{i\uparrow}n_{j\downarrow}\rangle$ for semions 
behave quantitatively similarly to those for fermions. This suggests that a
BCS-BEC crossover 
occurs in the semion superconductor. At 
$U=-5,-3,-1$ in Fig.~\ref{fig:rhoij}(a), $\rho(i,j)$ for semions is always 
smaller than that for fermions. Noting the absence of 
superconducting states in noninteracting systems of fermions, one may expect 
an absence of superconductivity for semions at $U=0$ as well.
This is, however, not necessarily correct. For both semions and fermions, the 
size of a pair becomes large as $U$ approaches zero and therefore the influence
of finite-size effects becomes non-negligible.
Much larger systems will be necessary to investigate the presence of
superconductivity for semions at $U=0$.

\section{Conclusion}
In this paper, we have constructed a formulation that allows for the Hubbard 
model of spinful anyons with any values of the on-site interaction. 
Virtual splitting of sites, which fixes the way opposing spins pass each other on a site, allows for double-spin occupancy.
Using this model, we have 
investigated the emergence of superconductivity for interacting semions. 
Off-diagonal long range order is 
numerically confirmed in the strong interaction regime. Our numerical results 
also suggest that a BCS-BEC crossover 
occurs in semionic systems.
Recently, density-dependent gauge potentials have been realized in cold
atoms~\cite{Clark18,Gorg19,Schweizer19,Lienhard20}. 
We believe that our findings will be useful for further explorations in the
physics of semions.

\begin{acknowledgments}
We wish to thank J. K. Jain for many fruitful discussions and many useful 
comments. We also thank G. J. Sreejith, Y. Kuno, and A. Sharma for 
helpful discussions.
K.K. thanks JSPS for support from Overseas Research Fellowship.
J.S. was supported in part by the U. S. Department of Energy, Office of Basic 
Energy Sciences, under Grant no. DE-SC-0005042.
We acknowledge Advanced CyberInfrastructure computational resources provided by
The Institute for CyberScience at The Pennsylvania State University. 
\end{acknowledgments}

\appendix
\section{Hopping phases and virtual splitting}
\label{appx:check}
\begin{figure}[]
 \begin{center}
  \includegraphics[width=\columnwidth]{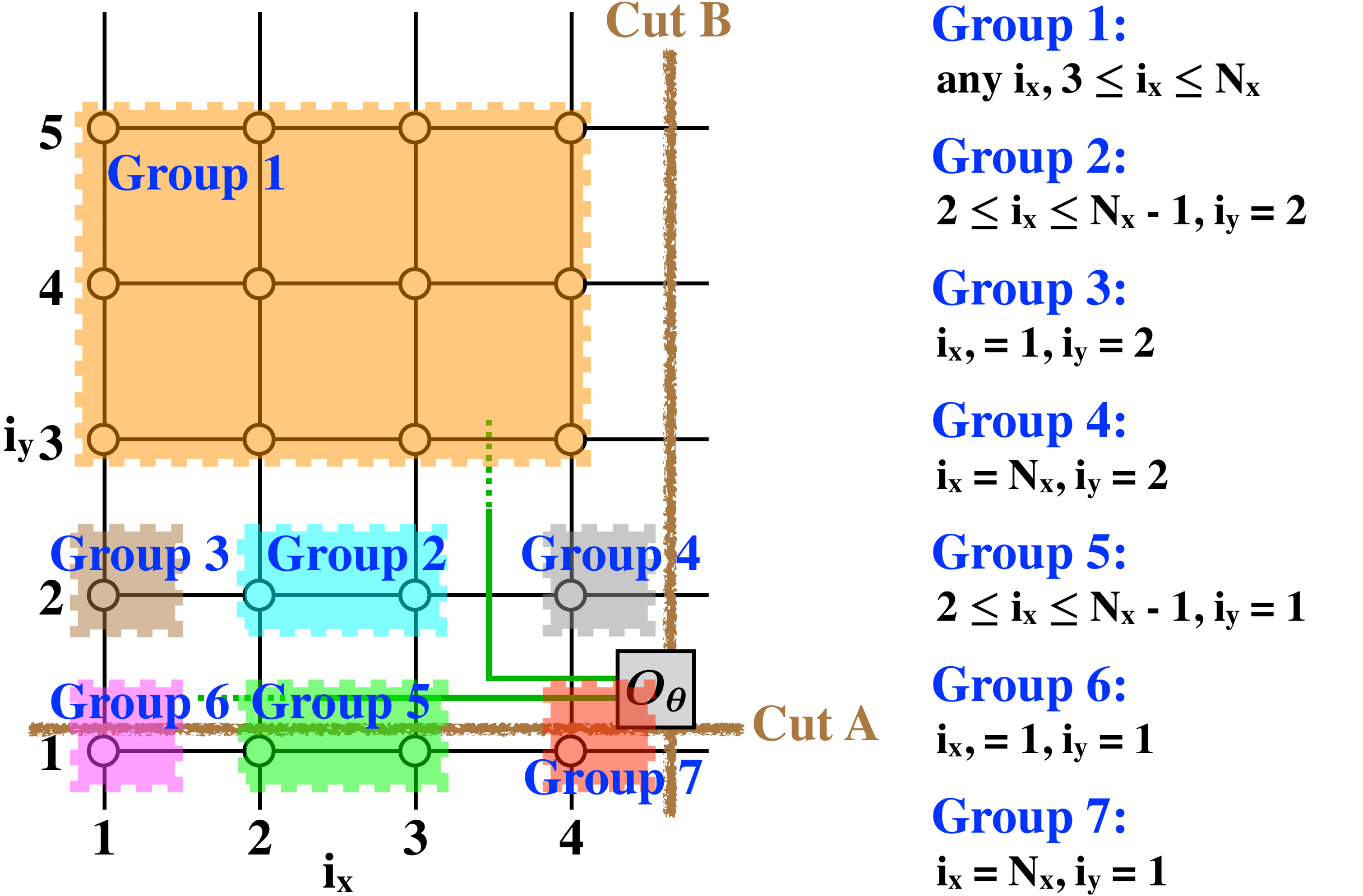}
 \end{center}
 \caption{
 Sketch of a $4\times5$ lattice. The sites are divided into seven groups.
 }
 \label{fig:7groups}
\end{figure}
\begin{figure*}[b]
 \begin{center}
  \includegraphics[width=2\columnwidth]{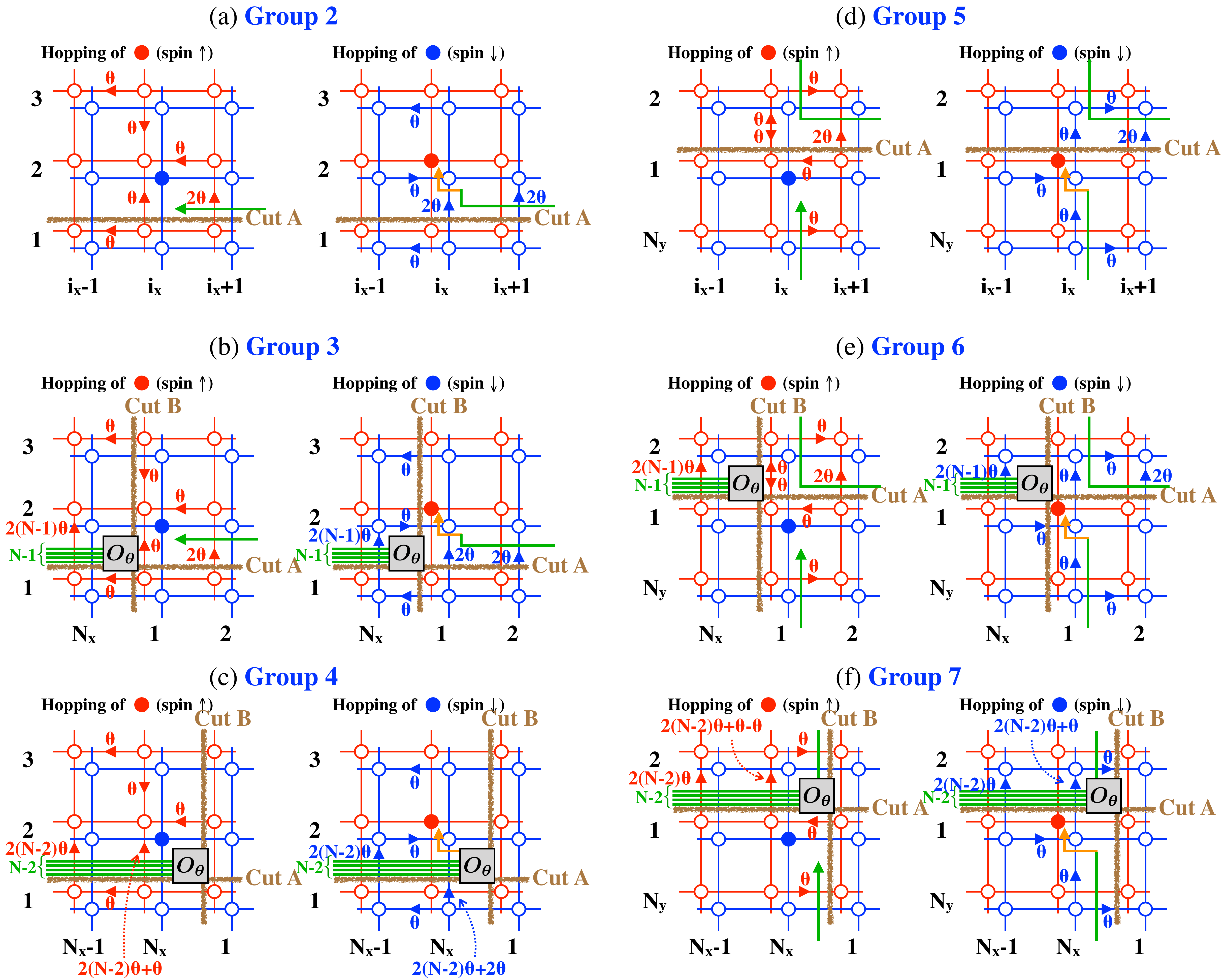}
 \end{center}
 \caption{
 The same as Fig.~\ref{fig:check1} but for the other groups shown in 
 Fig.~\ref{fig:7groups}. (Figure~\ref{fig:check1} corresponds to Group 1.)
 We here do not show the phase factors given by $W_{ij\alpha}$, i.e., the 
 plaquette with $O_\theta$ has the additional flux $-2\theta$ 
 [see Eq.~\eqref{eq:WWWW}] other than the sum of the written phases.
 }
 \label{fig:check2-7}
\end{figure*}
In this section, we show that the rules of $\theta_{ij\alpha}$ and 
$W_{ij\alpha}$ defined in the main text describe virtual splitting. As 
shown in Fig.~\ref{fig:7groups}, we first divide the sites of the system into 
seven groups and then calculate the hopping phases for each case 
in Fig.~\ref{fig:check2-7} as we did in
Fig.~\ref{fig:check1}. For all cases, the sum of
the hopping phases along each plaquette matches with the virtual splitting.
For simplicity, we impose some conditions of the configuration of the other 
$N-2$ particles out of the figure without loss of generality 
[e.g. there are no particles at $i_x=N_x$ in Fig.~\ref{fig:check2-7}(b)].

\section{Effective Hamiltonian for fermions at $U\ll-1$}
\label{appx:Hfermions}
Let us rewrite the Hamiltonian in Eq.~\eqref{eq:ham_spinful} for $\theta/\pi=1$
(fermions) as $H=H_\text{kin}+H_\text{int}$ with
\begin{align}
 &H_\text{kin}=-t\sum_{\langle ij\rangle,\alpha}
 f^\dagger_{i\alpha}f_{j\alpha},\\
 &H_\text{int}=U\sum_i\hat{N}_{i\uparrow}\hat{N}_{i\downarrow},
\end{align}
where 
$f^\dagger_{i\alpha}$ is the fermion operator and 
$\hat{N}_{i\alpha}=f^\dagger_{i\alpha}f_{i\alpha}$.
To see the expression of 
$H'=PH_\text{kin}Q\frac{1}{E_0-H_\text{int}}QH_\text{kin}P$, we consider a 
two-body state:
\begin{align}
 &PH_\text{kin}\frac{1}{E_0-H_\text{int}}H_\text{kin}
 f^\dagger_{j\uparrow}f^\dagger_{j\downarrow}\ket{0}\non
 =&PH_\text{kin}\frac{-t}{U}\sum_i
 \left(
 f^\dagger_{i\uparrow}f^\dagger_{j\downarrow}
 -f^\dagger_{i\downarrow}f^\dagger_{j\uparrow}
 \right)
 \ket{0} \non 
 =&\frac{2t^2}{U}\sum_i
 \left(
 f^\dagger_{j\uparrow}f^\dagger_{j\downarrow}
 +f^\dagger_{i\uparrow}f^\dagger_{i\downarrow}
 \right)
 \ket{0},
\end{align}
where $\sum_i$ indicates the summation over the four nearest sites of 
$j$.
We have
$[f^\dagger_{i\uparrow}f^\dagger_{i\downarrow},f_{j\uparrow}f_{j\downarrow}]
=\left(1-\left(N_{i\uparrow}-N_{i\downarrow}\right)\right)\delta_{ij}$ and 
$P\left(N_{i\uparrow}-N_{i\downarrow}\right)P=0$. Thus, the effective 
Hamiltonian $H'$ is given by
\begin{align}
 H'=\frac{2t^2}{U}\sum_{\langle ij\rangle}b^\dagger_ib_j
 +\frac{8t^2}{U}\sum_{i}b^\dagger_ib_i,
\end{align}
where $b_i^\dagger$ is the creation operator for a hard-core boson.
The twisted boundary conditions for $H'$ are given by 
$(\eta'_x,\eta_y')=(2\eta_x,2\eta_y)$, where $(\eta_x,\eta_y)$ is the 
angles for fermions in the original Hamiltonian. The effective Hamiltonian for semions could be derived using a similar argument, albeit with a modification of the boundary conditions, as discussed in the main text.

\section{BCS-BEC crossover}
\label{appx:BCS}
\begin{figure}[]
 \begin{center}
  \includegraphics[width=\columnwidth]{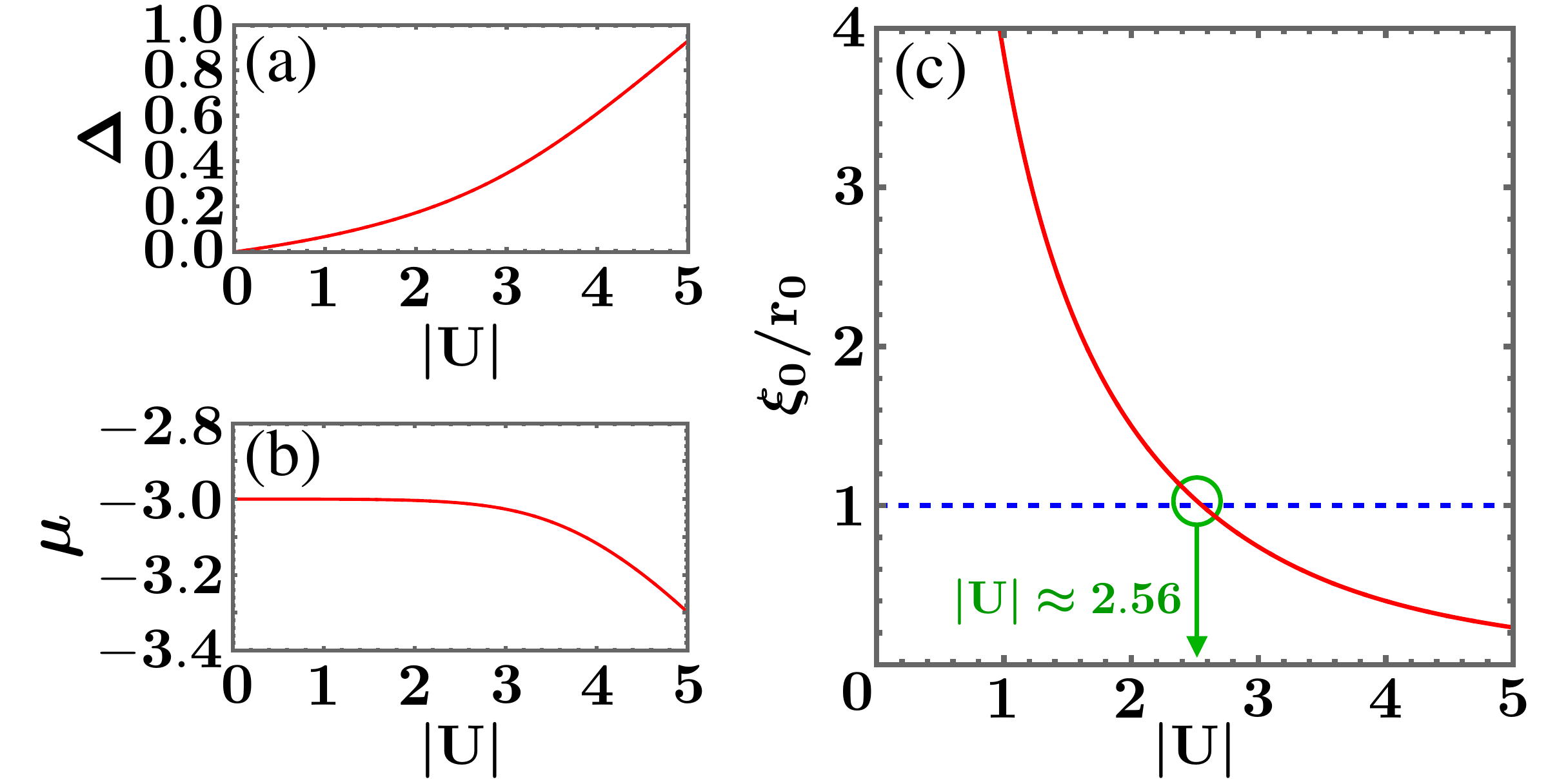}
 \end{center}
 \caption{
 (a) Order parameter $\Delta$, (b) chemical potential $\mu$, and (c) the ratio 
 of the size of the Cooper pair $\xi_0$ to the inter-particle distance $r_0$ as
 functions of the Hubbard interaction $|U|$. (c) The dashed line represents 
 $\xi_0=r_0$. The green circle at $|U|\approx2.56$ indicates the intersection 
 point.
 }
 \label{fig:BCS-BEC}
\end{figure}
The Hamiltonian in Eq.~\eqref{eq:ham_spinful} for $\theta/\pi=1$ reproduces the
standard Hubbard model of fermions. In this section, We calculate the value of 
$U$ for which the BCS-BEC crossover occurs in the fermionic system.

Assuming $U<0$, we rewrite the Hamiltonian in the reciprocal space as 
$H=H_\text{kin}+V$ with
\begin{align}
 &H_\text{kin}
 =\sum_{\bm{k}}\epsilon_{\bm{k}}\hat{N}_{\bm{k}},\\
 &V=\frac{-|U|}{N_xN_y}\sum_{\bm{k}_1,\bm{k}_2,\bm{k}_1',\bm{k}_2'}
 \delta_{\bm{k}_1+\bm{k}_2,\bm{k}_1'+\bm{k}_2'}
 f^\dagger_{\bm{k}_1\uparrow}f^\dagger_{\bm{k}_2\downarrow}
 f_{\bm{k}_2'\downarrow}f_{\bm{k}_1'\uparrow},
\end{align}
where $f^\dagger_{\bm{k}\alpha}$ is the fermionic operator,
$\hat{N}_{\bm{k}}=f^\dagger_{\bm{k}\uparrow}f_{\bm{k}\uparrow}
 +f^\dagger_{\bm{k}\downarrow}f_{\bm{k}\downarrow}$, and
$\epsilon_{\bm{k}}=-2t\big(\cos(k_xa)+\cos(k_ya)\big)$ with $t=1$. Here we 
explicitly write the lattice constant $a$. The number of sites is 
$N_x\times N_y$. With the condition $\bm{k}_1+\bm{k}_2=0$, we modify the 
interaction as
\begin{align}
 V=\sum_{\bm{k},\bm{k}'}
 V(\bm{k},\bm{k}')
 f^\dagger_{\bm{k}\uparrow}f^\dagger_{-\bm{k}\downarrow}
 f_{-\bm{k}'\downarrow}f_{\bm{k}'\uparrow},
\end{align}
where $V(\bm{k},\bm{k}')=-|U|/(N_xN_y)$. Now we consider the gap equation at 
$T=0$:
\begin{align}
 \Delta_{\bm{k}}
 =-\sum_{\bm{k}'}V(\bm{k},\bm{k}')\frac{\Delta_{\bm{k}'}}{2E_{\bm{k}'}},
 \label{eq:gapEq1}
\end{align}
where $\Delta_{\bm{k}}$ is the order parameter and
$E_{\bm{k}}=\sqrt{(\epsilon_{\bm{k}}-\mu)^2+\Delta_{\bm{k}}^2}$. Since the 
particle number is fixed in our model, we choose the chemical potential $\mu$ 
such that 
\begin{align}
 \sum_{\bm{k}}\langle \hat{N}_{\bm{k}}\rangle
 =\sum_{\bm{k}}(1-(\epsilon_{\bm{k}}-\mu)/E_{\bm{k}})=N.
 \label{eq:chemical}
\end{align}
Now we assume 
$\Delta_{\bm{k}}=\text{const.}\equiv\Delta$. Setting the system parameters as
$N_x=N_y=6$ and $N=6$ and solving Eqs.~\eqref{eq:gapEq1} and 
\eqref{eq:chemical} simultaneously, we generate Figs.~\ref{fig:BCS-BEC}(a) and 
(b) that plot the solutions of $\Delta$ and $\mu$ as functions of $U$. 

One expects that the BCS-BEC crossover occurs when the size of the Cooper pair 
is comparable with the inter-particle distance $r_0=a\sqrt{N_xN_y/N}$. 
In the continuum limit, the size of the Cooper pair is characterized by 
\begin{align}
 \xi_0=\frac{\hbar v_F}{\pi\Delta}
 =\frac{\hbar}{\pi\Delta}\sqrt{\frac{2\epsilon_F}{m}}.
\end{align}
Substituting $\hbar^2/(2m)=ta^2$ and $\epsilon_F=\mu+4t$ (we here measure the 
kinetic energy relative to the bottom of the band), we have 
$\xi_0=2a\sqrt{t(\mu+4t)}/(\pi\Delta)$. Figure~\ref{fig:BCS-BEC}(c) plots the 
ratio $\xi_0/r_0$ as a function of $U$. One obtains $\xi_0=r_0$ at 
$|U|\approx2.56$.

\bibliography{biblio_fqhe}

\begin{thebibliography}{50}
\bibitem[{com()}]{com}
\bibinfo{note}{This feature is consistent with the relations
  $\tau_{i+1}=e^{-i2\theta}\tau_i$ and $\rho_{i+1}=e^{i2\theta}\rho_i$ in the
  problem of spinless anyons~\cite{Einarsson90,Hatsugai91}.}
\end{thebibliography}
\bibliographystyle{apsrev}
------------------------------

\end{document}